\documentclass[twocolumn,showpacs,preprintnumbers,amsmath,amssymb,floatfix]{revtex4}
\usepackage{graphicx}% Include figure files
\usepackage{dcolumn}% Align table columns on decimal point
\usepackage{bm}% bold math
\usepackage{fancybox}
\newcommand{\ocite}[1]{\cite{#1}}

\def\H0{H^0}

\def\connect#1{\leavevmode{\setbox1=\hbox{#1}\copy1%
\raise .2\ht1 \vbox{\moveleft \wd1\vbox{\hrule width \wd1 height .5pt depth 0pt}}%
}}

\def\ftn[#1]{\rlap{\footnotemark[#1]}}

\def\EMAX{  E^{\rm APW}_{\rm MAX} }
\def\EMAXm{ E^{\rm rmesh}_{\rm MAX} }

\begin{document}

\title{A fusion of the LAPW and the LMTO methods:
the augmented plane wave plus muffin-tin orbital (PMT) method
}
\author{Takao Kotani}
\affiliation{School of Materials, Arizona State University, Tempe, AZ, 85284}
\author{Mark van Schilfgaarde}
\affiliation{School of Materials, Arizona State University, Tempe, AZ, 85284}
%draft
\date{\today}

\begin{abstract}
We present a new full-potential method 
to solve the one-body problem, for example, in the local density approximation. 
The method uses the augmented plane waves (APWs) and the generalized
muffin-tin orbitals (MTOs) together as basis sets to represent 
the eigenfunctions. Since the MTOs can efficiently describe 
localized orbitals, e.g, transition metal 3$d$ orbitals, 
the total energy convergence with basis size is drastically improved
in comparison with the linearized APW method. Required parameters 
to specify MTOs are given by atomic calculations in advance. 
Thus the robustness, reliability, easy-of-use, and efficiency 
at this method can be superior to the linearized APW and MTO methods.
We show how it works in typical examples, Cu, Fe, Li, SrTiO$_3$, and GaAs.
%Parameter sets to specify MTOs can be prepared automatically by atomic calculations 
%in advance. 
%Therefore, the robustness, reliability, easy-of-use, 
%efficiency, at this method can be superior to the linearized APW and MTO methods. 

\end{abstract}

\pacs{71.15.Ap, 71.15.Fv 71.15.-m}

\maketitle
%%%%%%%%%%%% Intro
There are several kinds of all-electron full potential (FP) methods
to obtain numerically-accurate solutions 
in the local density approximation 
to the density functional theory \cite{ks65}.
Among such FP methods, most popular ones are the
linearized augmented plane wave (LAPW) method,
and the projector augmented-wave (PAW) method 
\cite{Andersen75,soler89,PAW,kresse99}.
They both use plane waves (PWs) to expand the
eigenfunctions in the interstitial regions.
However, PWs do not efficiently
describe the localized character of eigenfunctions  
just around the muffin-tins (MTs). For example, oxygen $2p$ 
(denoted as O($2p$) below) and transition metal's
3$d$ orbitals are well localized and atomic-like
even in solids, thus we need to superpose many PWs 
to represent these orbitals. 
For example, as shown in Ref.\onlinecite{liu94} (and below), 
the energy cutoff for the augmented PW(APW) $\EMAX >$ 15Ry 
is required in fcc Cu to obtain $\sim$ 1 mRy convergence 
for total energy in LAPW.
%This makes the calculation very inefficient; a typical case
%is the Cu impunity problem in bulk Si. Then we
%inevitably need to use high energy cutoff
%for PW $E^{\rm PW}_{\rm cut}$ just due to Cu.
In contrast, such orbitals can be rather effectively represented
by localized basis in real space. In fact, it is 
already accomplished in the linearized muffin-tin orbital (LMTO) method, 
which differs from the LAPW method in 
that envelope functions consists of
atomic-like localized orbitals
\cite{nfpmanual,lmfchap} instead of PWs. Such a localized 
augmented waves are called as MT orbital (MTO). 

To circumvent the inefficiency in the LAPW method,
we have implemented a new method named
as linearized augmented plane wave plus muffin-tin 
orbital (PMT) method. The PMT method includes
not only the APWs but also MTOs in its basis set.
Our implementation becomes LAPW in the no MTO limit. 
As we show later, we can very effectively 
reduce the number of basis set by including MTOs;
we see the rapid convergence of the total energy
as a function of the number of APWs (or energy cutoff $\EMAX$).
As far as we tested, APWs with $\EMAX \sim 5$ Ry 
in addition to minimum MTOs will be reasonably good enough for usual applications; 
e.g. for $<$1 mRy convergence of total energy for Cu. 
Even in comparison with the LMTO method of Ref.\cite{lmfchap},
the PMT method is quite advantageous in its simplicity. 
The parameters to specify these minimum MTOs 
($E_{\rm H}$ and $R_{\rm H}$ for each $l$ channel. See next paragraph.) 
are automatically determined by atomic calculations in advance. 
This is a great advantage practically because
optimization of these parameters is a highly non-linear problem \cite{lmfchap}
which makes the LMTO difficult to use.
%Further, we do not need to enlarge the basis with floating orbitals.
%to check the total energy convergence and to obtain accurate high energy bands.
Thus the PMT method can satisfy the requirements 
for latest first-principle methods, reliability, speed, easy-of-use, 
and robustness very well.
%We can safely apply the PMT method to systems filled
%with vacuum regions very efficiently (note that usual LAPW/PAW
%requires high $\EMAX \sim 15 Ry$ \cite{liu94} even when a Cu in 
%empty supercell).
In what follows, we explain points in our method, 
and then we show how it works.

\begin{figure*}[htbp]
\centering
\includegraphics[angle=0,scale=.7]{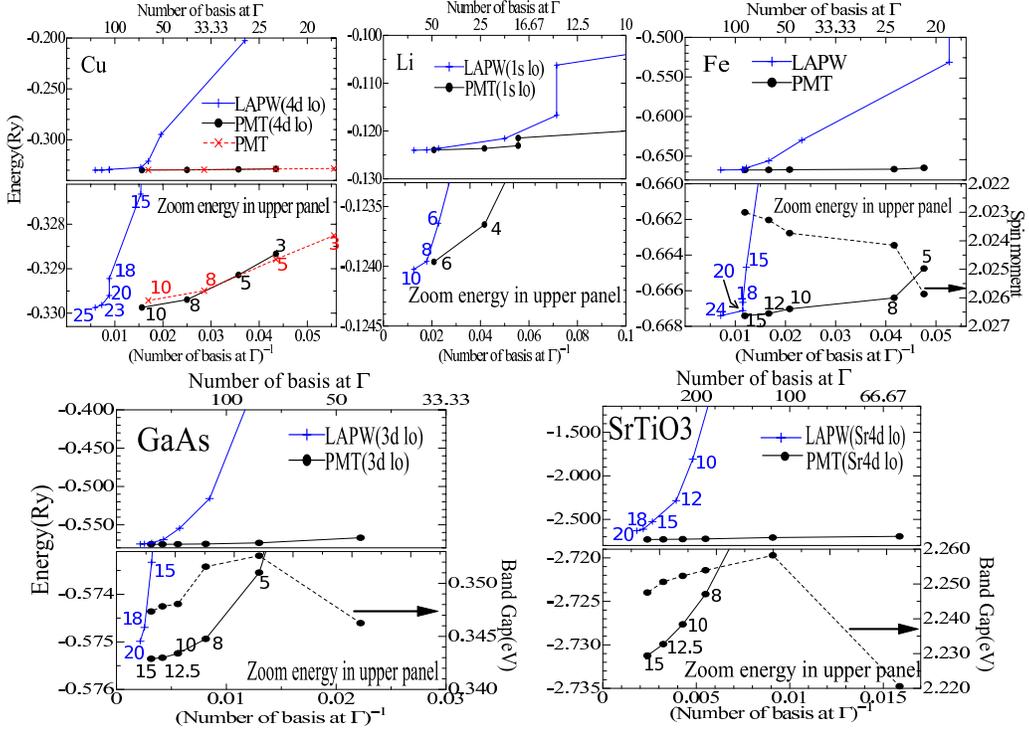}
\caption[]{(color online) Total energy is plotted as a function of
(number of basis at $\Gamma$ point $N^{\Gamma}_{\rm b}$)$^{-1}$.
Lower panels for each material zoom up upper panels in energy
(in SrTiO$_3$, data points in LAPW are out of the scale in lower panel).
We also plot spin moment ($\mu_{\rm B}$) for Fe,
and the band gaps for GaAs and SrTiO$_3$ by dotted lines, referring to right y-axis. 
The absolute values of energies are ambiguous
(See text in Table~\ref{tab:latt}). 
$\EMAX$(Ry) are shown next to each data point.
``$4d$ lo'' means we treat $4d$ as local orbital.
PMT denotes our new method with APW+MTO, where we use minimum MTO basis; 
number of MTOs (including lo) are 
14(Cu),5(Li),9(Fe),18(GaAs), and 33(SrTiO$_3$).
%Cu($4d$) is treated as lo for the case noted as ``PMT($4d$ lo)'', 
%though it is located high above the Fermi energy 
%(see the Fig.\ref{fig:cuband}).
Number of k point in 1st Brillouin zone (BZ)
are 12$^3$ for Cu and Li, 20$^3$ for Fe, and 6$^3$ for GaAs and SrTiO$_3$.
Lattice constants are not necessarily
at their total energy minima (e.g., 6.8a.u for Cu).
See Table~\ref{tab:latt} for exchange-correlation used.}
\label{fig:scaling}
\end{figure*}

%%%%%%%%%%%% Method
We adapted a variant of the LMTO method developed 
by Methfessel, van Schilfgaarde, and their collaborators 
\cite{nfpmanual,lmfchap}. 
%As the envelope functions 
%instead of PW in LAPW, 
This method uses 
the ``smooth Hankel function'' invented by
Methfessel as a modification of the usual Hankel functions 
so as to mimic the atomic orbitals \cite{lmfchap,Bott98}. 
It contains an additional parameter called as
the smoothing radius $R_{\rm H}$ in addition to the usual 
energy parameter $E_{\rm H}$ which specifies the damping behavior.
By choosing $R_{\rm H}$, we can control bending of the function just 
outside of MT. By using this degree of freedom, 
we can reproduce eigenvalues of atoms very well even 
if we substitute the eigenfunction outside of MT with
such a smooth Hankel function. This is an important 
feature of the function in comparison with others like 
Gaussians,which are not directly fit to the atomic orbitals.
Analytic properties of the smooth Hankel are analyzed 
in detail by Bott et al, and all the required operations, 
e.g, Bloch sum, to perform electronic structure calculations 
are well established \cite{Bott98}.
The augmentation procedure
requires the one-center expansion of the envelops functions.
In our method, the one-center expansion is 
given as the linear combinations of the generalized Gaussian
(Gaussians $\times$ polynomials), 
where the expansion coefficients are determined by
a projection procedure as described in Sec.XII in Ref.~\onlinecite{Bott98}.
Then the generalized Gaussians in each angular momentum $lm$-channel 
are replaced by the linear combinations of a radial function 
$\phi_l$ and its energy derivative $\dot{\phi}_l$ in the 
usual manner of augmentation \cite{lmfchap}.
When we use high energy cutoff $\EMAX \sim 15$Ry, we needed to use 
$\sim 15$ generalized Gaussians for the one-center expansion;
however, $\sim 5$ is good enough for practical usage with $\EMAX=5$ Ry.

Another key point in our method is the
smoothed charge density treatment introduced by
Soler and Williams \cite{soler89} to the LAPW method. 
The charge density is divided into the smooth part, 
the true density on MTs, and the counterpart on MTs. 
The smooth part is tabulated on regular uniform mesh in real space, 
and the others are tabulated on radial mesh in the spherical 
harmonics expansion in each MT sphere.
The PAW methods \cite{PAW,kresse99} also use this treatment.
It enables low-angular momentum $l$ cutoff 
for augmentation and makes the calculation very efficient.
As for the regular mesh in real space,
we usually need to use the spatial resolution
corresponding to the cutoff energy $\EMAXm= 10 \sim 15$ Ry,
which is determined so as to reproduce the smooth Hankel 
function well in real space. 
%We do not need to care
%$\EMAX$ so much when we choose $\EMAXm$ because only 
%the most localized smooth Hankels around an atom 
%should require the highest spatial resolution. 
Note that we still have some inefficiency, e.g. 
an atom in a large supercell requires a fine mesh everywhere
only in order to describe the density  
around the atom. This problem is common to
any method which uses an uniform mesh for density. 
%This is a problem which should be solved (not treated here).

Though the LMTO formalism shown in \cite{nfpmanual,lmfchap} is 
intended for such MTOs constructed 
from the smooth Hankel functions, it is essentially 
applicable to any kind of envelope functions.
%to the cases including PWs as members of the envelope 
%functions. 
As we explained above, our formalism is not
so different from LAPW/PAW formalisms shown in \cite{kresse99,PAW} 
except in the augmentation(projection) procedure.
The atomic forces are calculated \cite{lmfchap,molforce}
in the same manner as in PAW \cite{kresse99}.
For deep cores, we usually use a frozen core approximation
which allows the extension of core densities outside of MT
(but no augmentation) \cite{nfpmanual}. 
Further, we use the local orbital (lo) method \cite{lo} 
in some cases; for example, to treat 3$d$ semicore for Ga (denoted as Ga($3d$[lo]) below), 
or to reproduce high-lying bands for Cu by Cu($4d$[lo]).

%%% Result %%%%%%%%%%%%%%%%%%%%%%%%%%%%%%%%%%%%%%%%%%
\noindent {\it \bf Results:} 
In Fig.\ref{fig:scaling}, we plot the total energies 
as functions of the inverse of the number of basis at the $\Gamma$ point
$(N^{\Gamma}_{\rm b})^{-1}$ in order to observe its convergence as
$N^{\Gamma}_{\rm b} \to \infty$ (the number of basis is controlled
by $\EMAX$ as shown on Fig.\ref{fig:scaling} together).
Here we includes minimum MTOs whose parameters 
$E_{\rm H}$ and $R_{\rm H}$ are fixed by the atomic calculations 
(here ``minimum'' means only the atomic bound states).
The lo's are included as explained in Fig.\ref{fig:scaling}.

There is a problem of linear dependency in the basis set 
when we use large $\EMAX$. For example, in Li, we could not include
MTOs of Li($2s2p$) (Li($2s$) and Li($2p$)) 
as basis at $\EMAX >6$, because then they are well expanded by PWs.
This occurs also for other cases; when we use $\EMAX$
high enough to expand a MTO,
the rank of the overlap matrix of basis set is reduced by one. 
%This is consistent with supported by the fact that 
%the energy curve by PMT looks smoothly connected to LAPW results.
Thus possible ways to use large $\EMAX$ are: 
(1) keep only well localized MTOs which are not yet expanded by
given $\EMAX$, or 
(2) remove a subspace of basis through the diagonalization procedure
of the overlap matrix before solving the secular equation.
For (2), we need to introduce some threshold to
judge the linear dependency. This can cause an artificial discontinuity 
when changing lattice constants and so on. Thus (1) should be safer, 
but here we use the procedure (2) with careful check so that such 
discontinuity do not occur.
We use the number of basis after the procedure (2) for 
$N^{\Gamma}_{\rm b}$ to plot Fig.~\ref{fig:scaling}.
Even for LAPW cases, we applied the procedure (2), e.g, 
to the case for $\EMAX=$20Ry in SrTiO$_3$; 
then we reduce the dimension of Hamiltonian 
from 606 to $N^{\Gamma}_{\rm b}=550$ 
(data point at the left end of SrTiO$_3$ in Fig.\ref{fig:scaling}).

%All the energy at which
%we calculate $\phi,\dot{\phi}$ are the larger energy of 
%center of gravity of occupied states or thexxxx
As is seen in all the cases, the PMT method shows smooth 
and rapid convergence for the total energy  
at $(N^{\Gamma}_{\rm b})^{-1} \to 0$.
On the other hand, the convergence in LAPW
(no MTO limit in our PMT implementation) shows a 
characteristic feature; it is way off until it reaches 
to some $\EMAX$, e.g, $\sim 15$ Ry in Cu. 
This is consistent with previous calculations \cite{liu94}. 
$\EMAX \sim 15$ is required to reproduce the $3d$ localized orbitals.
We also show the convergence behaviors for band gap 
and magnetic moments by dotted lines (right y-axis); they are quite satisfactory.

\begin{figure}[htbp]
\centering
\includegraphics[angle=0,scale=.45]{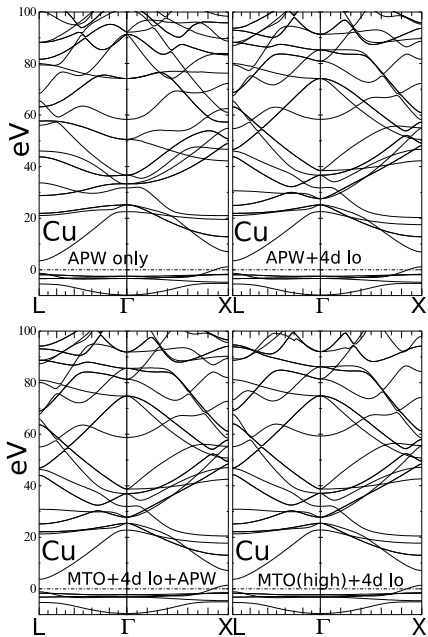}
\caption[]{
Energy bands for Cu up to 100 eV above the Fermi energy.
Top-left is for APW only without local orbitals.
Top-right: APW +$4d$[lo]. Bottom-left: APW plus MTO
(9 basis)+$4d$[lo]. Bottom-right: MTO(34 basis) +$4d$[lo].
All are converged for $\EMAX$.}
\label{fig:cuband}
\end{figure}

\begin{figure}[htbp]
\centering
\includegraphics[angle=-90,scale=.25]{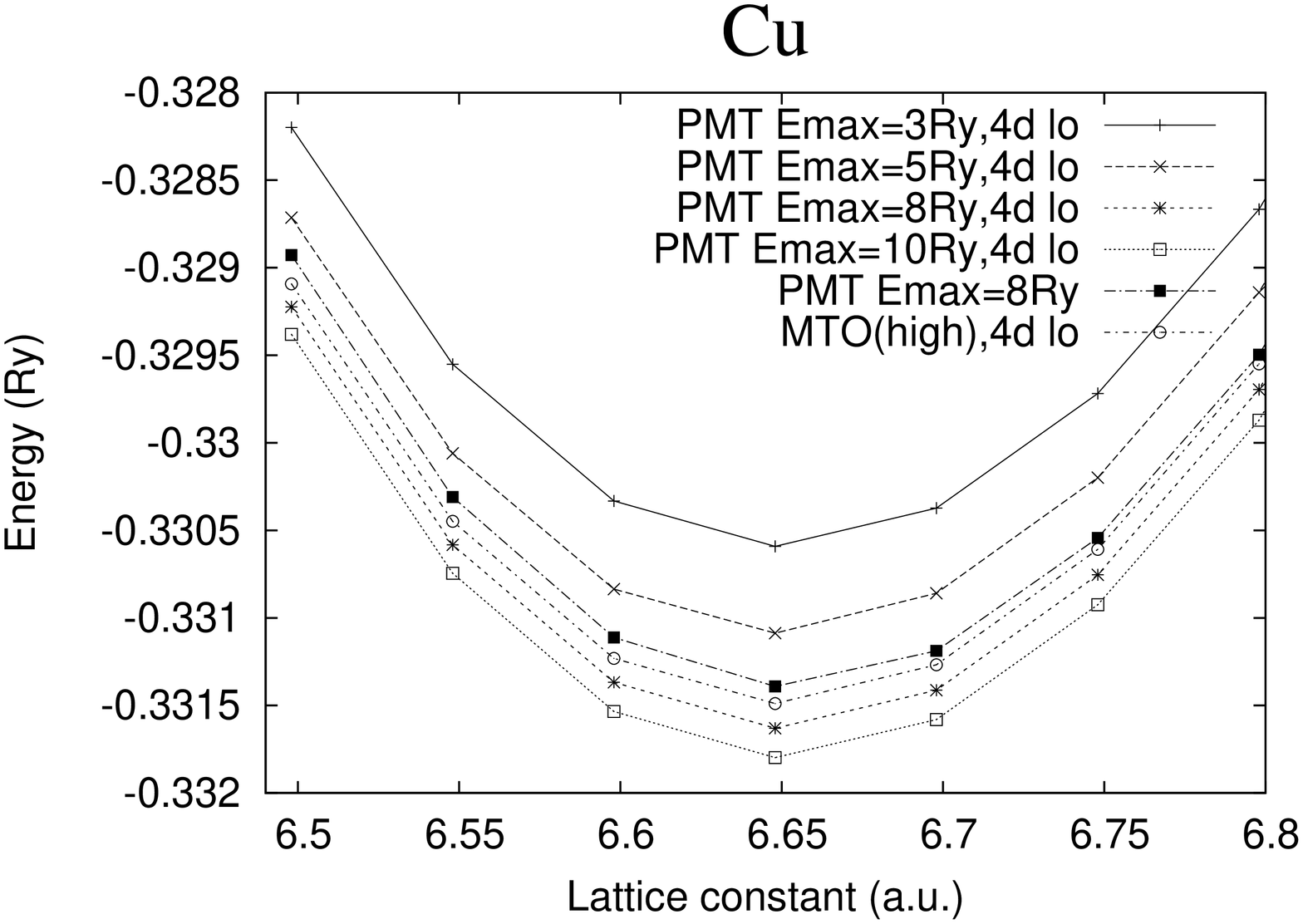}
\includegraphics[angle=-90,scale=.25]{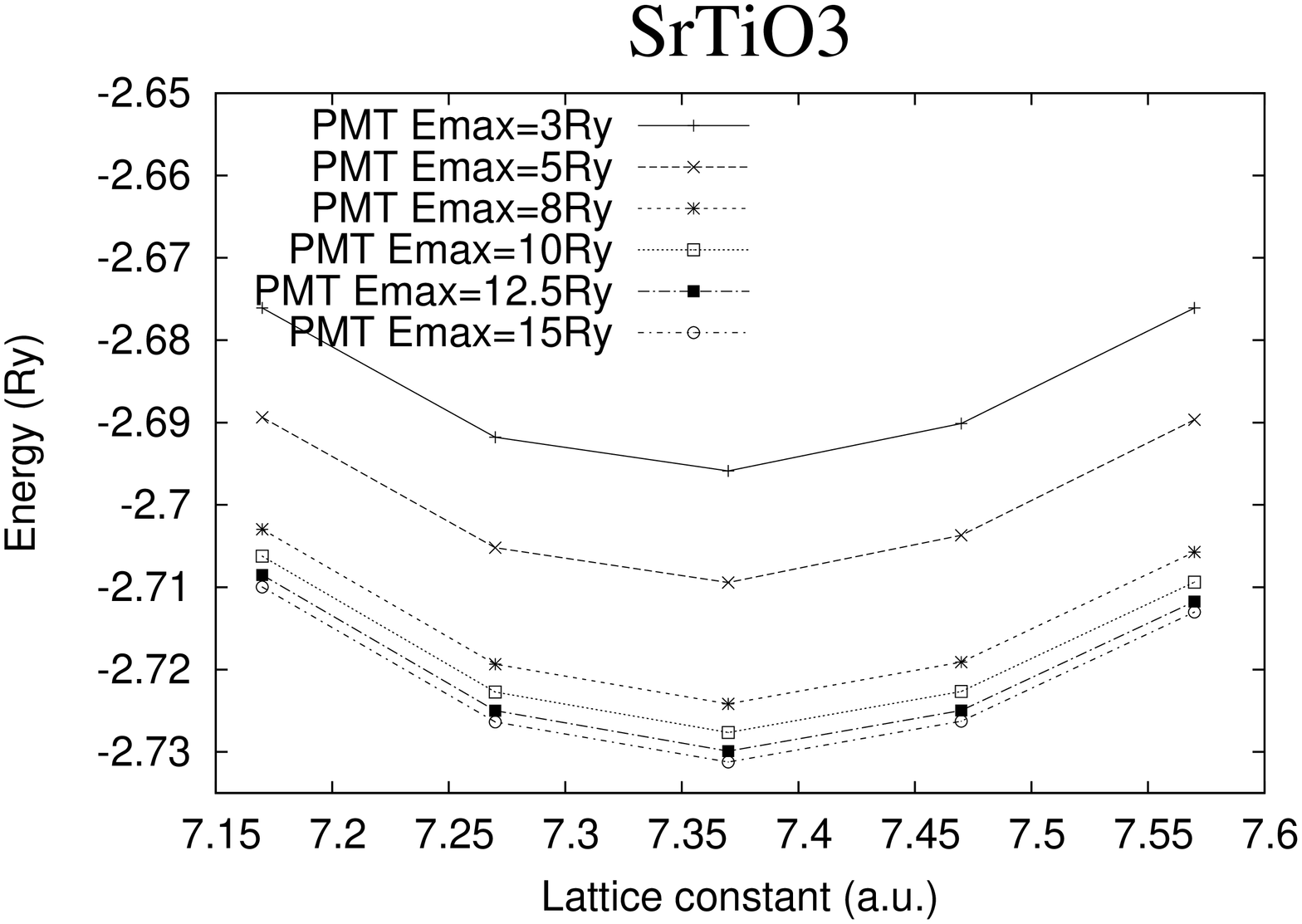}
\caption[]{Total energy v.s. lattice constant.
Labels ``Emax'' means $\EMAX$.
MTO(high) for Cu is explained in text.}
\label{fig:latt}
\end{figure}

Fig.\ref{fig:cuband} shows the energy bands up to 100eV above the
Fermi energy. Though the Cu panel in Fig.\ref{fig:scaling} shows 
the little effects of $4d$ local orbital 
to the total energy, it affects energy bands above $\sim30$eV. 
Calculations including $4d$[lo] gives good agreements each other. 
This means that we have no artificial bands (ghost bands). 
The MTO(high) panel is by the 
pure LMTO method where we use 34+5 basis ($spdfg+spd$+$4d$[lo]).
With some careful choice of $E_{\rm H}$ and $R_{\rm H}$,
the LMTO method can be very efficient and accurate.

%%% A1 B4 C5 D6 E8 F11 
\begin{figure}[htbp]
\centering
\includegraphics[angle=0,scale=.3]{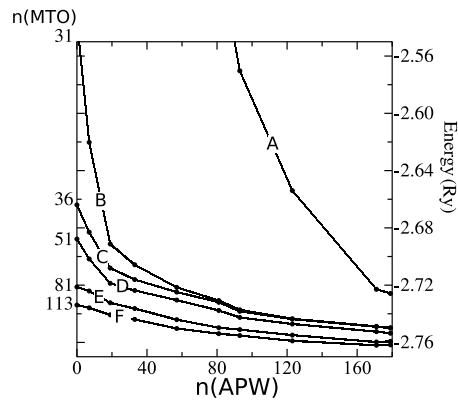}
\caption[]{Total energy for different type of MTO with 
changing the number of APWs n(APW) for SrTiO$_3$.
Each curves(A,B,C,D,E,F) corresponds to a type of MTO sets (see texts).
For example, curve C uses 36 MTOs (including local orbital). 
Thus curve C at n(APW)=80 means a calculation with n(MTO)+n(APW)=36+40=76 
basis. The converged energy $\sim -2.76$ Ry is a little different from
the Fig.\ref{fig:scaling} because this calculation is 
performed with coarse conditions for test purpose.}
\label{fig:srtio3mto}
\end{figure}

Fig.\ref{fig:latt} shows the total energy as
function of the lattice constant. All lines looks very parallel. 
This shows that PMT do not include some 
systematic errors. This is true also for other materials (not shown). 
Table \ref{tab:latt} shows the obtained lattice constants
and related parameters. We also showed values for $\EMAX$= 5Ry.
Even though we still need other extensive tests, we think that 
such low cutoff is reliable enough for practical applications.
For $\EMAX$= 5Ry, we only need $\sim 25$ basis 
(when we do not include lo) for Cu.

\begin{table}
\caption{
\baselineskip 12pt
Calculated lattice constant, bulk modulus, and cohesive energies.
Values are by the PMT method with $\EMAX=15$Ry 
($\EMAX=6$Ry for Li). Values in parenthesis are with $\EMAX=5$Ry. 
We show values from other literatures together.
We used the Barth-Hedin exchange correlation \cite{vbarth72} 
except Fe where we use the VWN type \cite{vwn}.
Since we have not determined the total energies 
of each atoms exactly (we assume spherical atoms and
assume some electron configurations), the cohesive energies
(constant in energy axis in Fig.~\ref{fig:scaling} and Fig.~\ref{fig:latt})
are somehow ambiguous; especially for cases including transition metals. 
For comparison with other calculations, it will
be better to use total energies; add -3304.4345Ry(Cu), -14.7106Ry(Li)
,-2540.4767Ry(Fe), -8397.5970Ry(GaAs), and -8502.4637Ry(SrTiO$_3$)
to the cohesive energies.
}
\begin{tabular}{l|*3{@{\hspace*{0.5em}}l}}
\vbox{\vskip 2pt}
           &  Lattice constant. & Bulk Modulus &Cohesive \\
           & (a.u.)& (GPa) & energy(Ry)\\
\colrule
  fcc Cu      &  6.650(6.649) & 188(187)   & -.332(-.331)   \\
  \ \  \ LAPW\ftn[1]  &  6.65 & 192 &    ---   \\
  \ \  \ expt.\ftn[1] &  6.81 & 131 &    ---   \\
  bcc Fe      &  5.209(5.208) & 258(259)   & -.667(-.665)   \\
  \ \  \ PAW\ftn[2]  &  5.20  & 247  &  ---  \\
  \ \  \ LAPW\ftn[3]  & 5.210 & 245 &    ---   \\
  \ \  \ expt.\ftn[3] &  5.417 & 172 &    ---   \\
  bcc Li      &  6.347(6.341) & 15.3(15.5) & -.124(-.123) \ftn[7] \\
  \ \  \ PAW\ftn[2]  &  6.355  & 15.0  &  -.149  \\
%  \ \  \ APW\ftn[6]  &  6.38  & 14.8   &  -.149  \\
   SrTiO$_3$          &  7.367(7.360) & 220(225)   & -2.731(-2.709) \\
  \ \  \ PP \ftn[4]   &  7.31  & 203 &    ---   \\
  \ \  \ expt.\ftn[5] &  7.39  & 184 &    ---   \\
   GaAs    &  10.61(10.61) & 74.9(75.0)   & -.576(-.574)   \\
  \ \  \ LAPW\ftn[5]  & 10.62  & 74 &   -.587   \\
  \ \  \ expt.\ftn[5] &  10.68 & 76 &   -.479   
\end{tabular}
\footnotetext[1]{Ref.~\ocite{khein95}}
\footnotetext[2]{Ref.~\ocite{kresse99}}
\footnotetext[3]{Ref.~\ocite{Stixrude94}}
\footnotetext[4]{Ref.~\ocite{liborio05}. PP denotes 
a pseudopotential method. }
\footnotetext[5]{Ref.~\ocite{filippi94}}
\footnotetext[7]{The difference from PAW may be because it uses the 
non-polarized Li atom as reference. If we do so, we have -.150(-.150).}
\label{tab:latt}
\end{table}
%\footnotetext[6]{PRB42 11637}

%%% A=1 B=4 C=5 D=6 E=8 F=11 
Fig.\ref{fig:srtio3mto} shows the total energies
with different MTO sets combined with different numbers of APWs.
Curve "A" includes just MTOs for O($2s2p$), 
sufficient for a crude representation of the valence bands. 
%(no orbitals are available for conduction bands). 
Also included are Sr($4p$[lo]) and Ti($3p$[lo]). 
%as they are too extended to be approximated as core states. 
A large number of APWs is needed to get a good total energy: 
$\sim 150$ APWs are needed to converge energy to 
within $\sim$50 mRy of the converged result. 
%(Had the O($2s2p$) are not been present, 
%many more APWs would have been required.)
"B" corresponds to an extreme tight-binding basis, 
consisting of Sr($5s5p$) and Ti($4s4p4d$) in addition to ``A''.
(Note that the conduction band is mainly Ti($3d$), and O($2s2p$)).
The total energy of the MTO basis alone (no APWs) is rather crude 
--- more than 200 mRy underbound. However, only 25 orbitals 
(plus 6 for lo's) are included in this basis. 
The energy drops rapidly as low-energy APWs are included: 
adding $\sim$40 APWs is sufficient to converge energy to $\sim$50 mRy. 
As more APWs are added, the gain in energy becomes 
more gradual; convergence is very slow for large $\EMAX$. 
"C" differs from "B" only in that Sr($3d$) orbital was added. 
With the addition of these 5 orbitals, the MTO-only basis is already rather 
reasonable. This would be the smallest acceptable MTO-basis. 
As in the "B" basis, there is initially a rapid gain in energy 
as the first few APWs are added, followed by a progressively slower 
gain in energy as more APWs are added. 
"D" is a standard LMTO minimum basis: $spd$ orbitals on all atoms. 
Comparing "C" or "E" to "A" shows that the MTO basis 
is vastly more efficient than the APW basis in converging the total energy. 
This is true until a minimum basis is reached. Beyond this point, 
the gain APWs and more MTOs improve the total energy with approximately 
the same efficiency, as the next tests show.
"E" is a standard LMTO larger basis: $spd+spd$ orbitals on Sr and Ti, 
and $spd+sp$ on O. Comparing "D" and 'F" shows that the efficiency 
of any one orbital added to to the standard MTO minimum basis is 
rather similar in the APW and MTO cases. Thus, increasing the MTO 
basis from 51 to 81 orbitals in the MTO basis lowers the energy by 33 mRy;
adding 33 APWs to the minimum basis ("D") lowers the energy by 36 mRy. 
%APWs, however, have the advantage that they are simpler to include.
"F" enlarges the MTO basis still more, with Sr: $spd+spd$, Ti: 
$spd+spd$, O: $spd+spd$. Also local orbitals are used to represent 
the high-lying states Ti($4d$[lo]) and O($3s$[lo],$3p$[lo]). 
For occupied states, these orbitals have little
effect for total energy as in the case of Cu.
% I am somehow skeptical about your converged value --- as far as I plotted,
% in Fig1, the total energy is still changing even above 500 basis.
% But I may use wrong MTO.

In conclusion, we have implemented the PMT method
whose basis set consists of the APWs together with the MTOs 
which are localized in real space.
This idea is consistent with the nature of the eigenfunctions 
in solids; they can be localized as atoms or extended as PW.
%From kinds of view points, the method is so advantageous
%than LAPW/PAW and LMTO. 
This method combines advantages of LAPW/PAW and LMTO. 
We have implemented force calculations, 
but no results are shown here. One of the advantage 
is in its flexibility. As an example, in order to treat 
Cu impurity in Si, it will be possible to choose very low 
$\EMAX$ just responsible for empty regions because
MTOs for Cu and MTOs for Si span its neighbors already very well. 
Convergence is easily checked by changing $\EMAX$ 
(much simpler than LMTO). In future, we can use the PMT method 
to make a natural division of the Kohn-Sham Hamiltonian 
into localized blocks and extended blocks, instead 
of the energy windows method for the maximally localized Wanner
functions \cite{souza01}.
The problem of large $\EMAXm$ should be solved to 
make PMT more efficient.

%\begin{acknowledgments}
This work was supported by ONR contract N00014-7-1-0479. 
We are also indebted to the Ira A. Fulton High Performance 
Computing Initiative.
%\end{acknowledgments}

\bibliography{lmto}

\begin{thebibliography}{18}
\expandafter\ifx\csname natexlab\endcsname\relax\def\natexlab#1{#1}\fi
\expandafter\ifx\csname bibnamefont\endcsname\relax
  \def\bibnamefont#1{#1}\fi
\expandafter\ifx\csname bibfnamefont\endcsname\relax
  \def\bibfnamefont#1{#1}\fi
\expandafter\ifx\csname citenamefont\endcsname\relax
  \def\citenamefont#1{#1}\fi
\expandafter\ifx\csname url\endcsname\relax
  \def\url#1{\texttt{#1}}\fi
\expandafter\ifx\csname urlprefix\endcsname\relax\def\urlprefix{URL }\fi
\providecommand{\bibinfo}[2]{#2}
\providecommand{\eprint}[2][]{\url{#2}}

\bibitem[{\citenamefont{Kohn and Sham}(1965)}]{ks65}
\bibinfo{author}{\bibfnamefont{W.}~\bibnamefont{Kohn}} \bibnamefont{and}
  \bibinfo{author}{\bibfnamefont{L.~J.} \bibnamefont{Sham}},
  \bibinfo{journal}{Phys. Rev.} \textbf{\bibinfo{volume}{140}},
  \bibinfo{pages}{A1133} (\bibinfo{year}{1965}).

\bibitem[{\citenamefont{Andersen}(1975)}]{Andersen75}
\bibinfo{author}{\bibfnamefont{O.~K.} \bibnamefont{Andersen}},
  \bibinfo{journal}{Phys. Rev. B} \textbf{\bibinfo{volume}{12}},
  \bibinfo{pages}{3060} (\bibinfo{year}{1975}).

\bibitem[{\citenamefont{Soler and Williams}(1989)}]{soler89}
\bibinfo{author}{\bibfnamefont{J.~M.} \bibnamefont{Soler}} \bibnamefont{and}
  \bibinfo{author}{\bibfnamefont{A.~R.} \bibnamefont{Williams}},
  \bibinfo{journal}{Phys. Rev. B} \textbf{\bibinfo{volume}{40}},
  \bibinfo{pages}{1560} (\bibinfo{year}{1989}).

\bibitem[{\citenamefont{Blochl}(1994)}]{PAW}
\bibinfo{author}{\bibfnamefont{P.~E.} \bibnamefont{Blochl}},
  \bibinfo{journal}{Phys. Rev. B} \textbf{\bibinfo{volume}{50}},
  \bibinfo{pages}{17953} (\bibinfo{year}{1994}).

\bibitem[{\citenamefont{Kresse and Joubert}(1999)}]{kresse99}
\bibinfo{author}{\bibfnamefont{G.}~\bibnamefont{Kresse}} \bibnamefont{and}
  \bibinfo{author}{\bibfnamefont{D.}~\bibnamefont{Joubert}},
  \bibinfo{journal}{Phys. Rev. B} \textbf{\bibinfo{volume}{59}},
  \bibinfo{pages}{1758} (\bibinfo{year}{1999}).

\bibitem[{\citenamefont{Liu et~al.}(1994)\citenamefont{Liu, Singh, and
  Krakauer}}]{liu94}
\bibinfo{author}{\bibfnamefont{A.~Y.} \bibnamefont{Liu}},
  \bibinfo{author}{\bibfnamefont{D.~J.} \bibnamefont{Singh}}, \bibnamefont{and}
  \bibinfo{author}{\bibfnamefont{H.}~\bibnamefont{Krakauer}},
  \bibinfo{journal}{Phys. Rev. B} \textbf{\bibinfo{volume}{49}},
  \bibinfo{pages}{17424} (\bibinfo{year}{1994}).

\bibitem[{\citenamefont{Mesfessel and van Schilfgaarde}()}]{nfpmanual}
\bibinfo{author}{\bibfnamefont{M.}~\bibnamefont{Mesfessel}} \bibnamefont{and}
  \bibinfo{author}{\bibfnamefont{M.}~\bibnamefont{van Schilfgaarde}},
  \bibinfo{note}{'NFP manual 1.01 Oct 10,1997'. NFP is previous to the current
  LMTO package lmf maintained by M. van Schilfgaarde.}

\bibitem[{\citenamefont{Methfessel et~al.}(2000)\citenamefont{Methfessel, van
  Schilfgaarde, and Casali}}]{lmfchap}
\bibinfo{author}{\bibfnamefont{M.}~\bibnamefont{Methfessel}},
  \bibinfo{author}{\bibfnamefont{M.}~\bibnamefont{van Schilfgaarde}},
  \bibnamefont{and} \bibinfo{author}{\bibfnamefont{R.~A.}
  \bibnamefont{Casali}}, in \emph{\bibinfo{booktitle}{Lecture Notes in
  Physics}}, edited by
  \bibinfo{editor}{\bibfnamefont{H.}~\bibnamefont{Dreysse}}
  (\bibinfo{publisher}{Springer-Verlag, Berlin}, \bibinfo{year}{2000}), vol.
  \bibinfo{volume}{535}.

\bibitem[{\citenamefont{Bott et~al.}(1998)\citenamefont{Bott, Methfessel,
  Krabs, and Schmidt}}]{Bott98}
\bibinfo{author}{\bibfnamefont{E.}~\bibnamefont{Bott}},
  \bibinfo{author}{\bibfnamefont{M.}~\bibnamefont{Methfessel}},
  \bibinfo{author}{\bibfnamefont{W.}~\bibnamefont{Krabs}}, \bibnamefont{and}
  \bibinfo{author}{\bibfnamefont{P.~C.} \bibnamefont{Schmidt}},
  \bibinfo{journal}{J. Math. Phys.} \textbf{\bibinfo{volume}{39}},
  \bibinfo{pages}{3393} (\bibinfo{year}{1998}).

\bibitem[{\citenamefont{Methfessel and van Schilfgaarde}(1993)}]{molforce}
\bibinfo{author}{\bibfnamefont{M.}~\bibnamefont{Methfessel}} \bibnamefont{and}
  \bibinfo{author}{\bibfnamefont{M.}~\bibnamefont{van Schilfgaarde}},
  \bibinfo{journal}{Phys. Rev. B} \textbf{\bibinfo{volume}{48}},
  \bibinfo{pages}{4937} (\bibinfo{year}{1993}).

\bibitem[{\citenamefont{E.~Sjostedt}(2000)}]{lo}
\bibinfo{author}{\bibfnamefont{D.~J.~S.} \bibnamefont{E.~Sjostedt},
  \bibfnamefont{L.~Nordstrom}}, \bibinfo{journal}{Solid State Communications}
  \textbf{\bibinfo{volume}{114}}, \bibinfo{pages}{15} (\bibinfo{year}{2000}).

\bibitem[{\citenamefont{von Barth and Hedin}(1972)}]{vbarth72}
\bibinfo{author}{\bibfnamefont{U.}~\bibnamefont{von Barth}} \bibnamefont{and}
  \bibinfo{author}{\bibfnamefont{L.}~\bibnamefont{Hedin}}, \bibinfo{journal}{J.
  Phys. C} \textbf{\bibinfo{volume}{5}}, \bibinfo{pages}{1692}
  (\bibinfo{year}{1972}).

\bibitem[{\citenamefont{S.~H.~Vosko and Nusair}(1980)}]{vwn}
\bibinfo{author}{\bibfnamefont{L.~W.} \bibnamefont{S.~H.~Vosko}}
  \bibnamefont{and} \bibinfo{author}{\bibfnamefont{M.}~\bibnamefont{Nusair}},
  \bibinfo{journal}{Can. J. Phys.} \textbf{\bibinfo{volume}{58}},
  \bibinfo{pages}{1200} (\bibinfo{year}{1980}).

\bibitem[{\citenamefont{Khein et~al.}(1995)\citenamefont{Khein, Singh, and
  Umrigar}}]{khein95}
\bibinfo{author}{\bibfnamefont{A.}~\bibnamefont{Khein}},
  \bibinfo{author}{\bibfnamefont{D.~J.} \bibnamefont{Singh}}, \bibnamefont{and}
  \bibinfo{author}{\bibfnamefont{C.~J.} \bibnamefont{Umrigar}},
  \bibinfo{journal}{Phys. Rev. B} \textbf{\bibinfo{volume}{51}},
  \bibinfo{pages}{4105} (\bibinfo{year}{1995}).

\bibitem[{\citenamefont{Stixrude et~al.}(1994)\citenamefont{Stixrude, Cohen,
  and Singh}}]{Stixrude94}
\bibinfo{author}{\bibfnamefont{L.}~\bibnamefont{Stixrude}},
  \bibinfo{author}{\bibfnamefont{R.~E.} \bibnamefont{Cohen}}, \bibnamefont{and}
  \bibinfo{author}{\bibfnamefont{D.~J.} \bibnamefont{Singh}},
  \bibinfo{journal}{Phys. Rev. B} \textbf{\bibinfo{volume}{50}},
  \bibinfo{pages}{6442} (\bibinfo{year}{1994}).

\bibitem[{\citenamefont{Liborio et~al.}(2005)\citenamefont{Liborio, Sanchez,
  Paxton, and Finnis}}]{liborio05}
\bibinfo{author}{\bibfnamefont{L.~M.} \bibnamefont{Liborio}},
  \bibinfo{author}{\bibfnamefont{C.~G.} \bibnamefont{Sanchez}},
  \bibinfo{author}{\bibfnamefont{A.~T.} \bibnamefont{Paxton}},
  \bibnamefont{and} \bibinfo{author}{\bibfnamefont{M.~W.}
  \bibnamefont{Finnis}}, \bibinfo{journal}{Journal of Physics: Condensed
  Matter} \textbf{\bibinfo{volume}{17}}, \bibinfo{pages}{L223}
  (\bibinfo{year}{2005}),
  \urlprefix\url{http://stacks.iop.org/0953-8984/17/L223}.

\bibitem[{\citenamefont{Filippi et~al.}(1994)\citenamefont{Filippi, Singh, and
  Umrigar}}]{filippi94}
\bibinfo{author}{\bibfnamefont{C.}~\bibnamefont{Filippi}},
  \bibinfo{author}{\bibfnamefont{D.~J.} \bibnamefont{Singh}}, \bibnamefont{and}
  \bibinfo{author}{\bibfnamefont{C.~J.} \bibnamefont{Umrigar}},
  \bibinfo{journal}{Phys. Rev. B} \textbf{\bibinfo{volume}{50}},
  \bibinfo{pages}{14947} (\bibinfo{year}{1994}).

\bibitem[{\citenamefont{Souza et~al.}(2001)\citenamefont{Souza, Marzari, and
  Vanderbilt}}]{souza01}
\bibinfo{author}{\bibfnamefont{I.}~\bibnamefont{Souza}},
  \bibinfo{author}{\bibfnamefont{N.}~\bibnamefont{Marzari}}, \bibnamefont{and}
  \bibinfo{author}{\bibfnamefont{D.}~\bibnamefont{Vanderbilt}},
  \bibinfo{journal}{Phys. Rev. B} \textbf{\bibinfo{volume}{65}},
  \bibinfo{pages}{035109} (\bibinfo{year}{2001}).

\end{thebibliography}

\end{document}